\documentclass[twocolumn,prl,amsmath,amssymb]{revtex4}% Physical Review Letters
 
\usepackage{graphicx}% Include figure files
\usepackage{dcolumn}% Align table columns on decimal point
\usepackage{bm}% bold math
\usepackage{feynmf} 
 
\newcommand{\bfr}{{\mathbf{r}}}

\newcommand{\dd}{{\mathrm{d}}}
\newcommand{\ee}{{\mathrm{e}}}
\newcommand{\tr}{{\mathrm{tr}}}

\newcommand{\baru}{{u^\dagger}}
\newcommand{\barY}{{\bar{Y}}}
\newcommand{\bara}{{\bar{\alpha}}}
\newcommand{\bareta}{{\bar{\eta}}}
\newcommand{\bfeta}{{\boldsymbol{\eta}}}
\newcommand{\bfbareta}{\bar{\boldsymbol{\eta}}}
\newcommand{\barpsi}{{\psi^\dagger}}
\newcommand{\barT}{{T^*}}                                                       
\newcommand{\inter}{{\mathrm{int}}}
\newcommand{\rhonew}{{\rho_\mathrm{new}}}

\renewcommand{\i}[1]{{}_{\scriptscriptstyle(#1)}}
 
\begin{document}
 
\preprint{APS/123-QED}
 
\title{A many-body approach to crystal field theory}
 
\author{Christian Brouder}
\affiliation{%
Laboratoire de Min\'eralogie Cristallographie, CNRS UMR 7590, 
Universit\'es Paris 6 et 7, IPGP, case 115, 4 place Jussieu, 
75252 Paris cedex 05, France.
}%
 
\date{\today}% It is always \today, today,
 
\begin{abstract}
A self-consistent many-body approach is proposed to build
a first-principles crystal field theory, where crystal field
parameters are calculated {\it{ab initio}}. Many-body theory
is used to write the energy of the interacting system as a function of the
density matrix of the noninteracting system. A variation
of the energy with respect to the density matrix gives
an effective Hamiltonian matrix that is diagonalized to determine
the density matrix providing the lowest energy. The equations
are written in terms of the quantum group of functional
derivatives with respect to external fermionic sources.
This approach contains the many-body theory of Green
functions as a special case and the usual crystal field
theory as a first approximation. Therefore, it is expected
to provide good results for strongly-interacting
electron systems.
\end{abstract}
 
\maketitle
 
The many-body theory of realistic quantum systems
uses two main methods: (i) the Green function
method that gives good results for extended states
such as metals and semiconductors
and (ii) the diagonalization method, typically used
in quantum chemistry, which describes
efficiently localized systems and the splitting of
degenerate states. The latter model is efficient
to calculate the magnetism
of rare earths and transition metal impurities
in insulators, as well as of their optical
and spectroscopic properties.
However, the second method leads to the diagonalization
of very large matrices for complex systems.
Thus, an intermediate approach is desirable,
where a small-size Hamiltonian
is set up using effective parameters.
The best-known example of this approach is
the crystal-field or ligand-field method, which
was extended by Kotani and coll. \cite{Kotani} to describe
the interaction between a localized and an extended
system. It gives good results for the calculation
of x-ray absorption and photoemission spectroscopies.

The main drawback of this method is its use
of parameters, which can be numerous in the
case of low symmetry.
The calculation of crystal field and 
Slater parameters in molecules and solids is a 
subject of continued interest
\cite{Richter,Novak,Fahnle,Shen,Brooks,Fahnle2,Colarieti,%
Bagus,Ogasawara,Ogasawara2001},
but they are not part of a general formalism
and double counting of interactions is 
difficult to control. Moreover,
the effect of the crystal field is not
taken into account self-consistently.

In this paper, we propose a first-principles
self-consistent approach of the effective Hamiltonian method,
based on the quantum field theory of correlated
systems \cite{BrouderKB0,BrouderKB}.

It was noticed a long time ago by Esterling and Lange
\cite{Esterling} that the many-body solution of
a degenerate or quasidegenerate system contains a
loophole: among the various (quasi) degenerate states of 
the noninteracting system, only one will evolve
towards the ground state of the interacting system,
and it must be known in advance to set up the
many-body calculation. 
Here we solve this problem by showing that 
the relevant state of the noninteracting system
can be obtained by diagonalizing an effective
Hamiltonian. Moreover, as shown in \cite{BrouderKB},
the requirement of self-consistency imposes
the use of density matrices instead of 
pure states. So, for a density matrix $\rho$,
we calculate the total energy of the interacting system
$E(\rho)$. To determine the
density matrix $\rho$ that minimizes the energy,
we have to diagonalize an effective matrix which
is determined by the quantum group approach
to many-body theory. Therefore, this
method unifies the diagonalization method and
the Green function method.

{\sl{Total energy of the system}} --
Standard many-body theory \cite{Gross} enables us to write the
total energy of an interacting system which evolves from 
a noninteracting system described by the density matrix $\rho$ as
\begin{eqnarray}
E(\rho)&=&
\tr\Big(\rho S^{-1} T\big(H(0)\ee^{-i\int_{-\infty}^\infty H^\inter(t)\dd t}\big) \Big),
\label{HSinv}
\end{eqnarray}
where $S$ is the S-matrix, $T$ is the time-ordering operator and
$H(t)=H_0(t)+H^\inter(t)$ is the total Hamiltonian in the
interaction representation, with
\begin{eqnarray*}
H_0(t) &=&
 \int \barpsi(t,\bfr) h_0(\bfr)
   \psi(t,\bfr) \dd\bfr,\\
H^\inter(t) &=&
 \frac{e^2}{2} \int
 \frac{\barpsi(t,\bfr)\barpsi(t,\bfr') \psi(t,\bfr')\psi(t,\bfr)}
      {|\bfr-\bfr'|}
 \dd\bfr \dd\bfr',
\end{eqnarray*}
where $h_0(\bfr)= -\Delta/2m + U_N(\bfr)$ is the
one-body Hamiltonian operator for an electron in
a nuclear potential $U_N(\bfr)$.

Within the functional derivative approach to
quantum field theory \cite{Chou,BrouderKB}, this expression can be
rewritten $E(\rho)=P(Z_\rho)$, where
\begin{eqnarray*}
P(Z_\rho) &=& \int\lim_{\bfr'\rightarrow \bfr} h_0(\bfr)
\frac{\delta^2 Z_\rho}{\delta \bareta^+(x)\delta \eta^+(y)} \dd\bfr
\\&&\hspace*{-5mm}+
\int \frac{e^2}{2|\bfr-\bfr'|} \frac{\delta^4 Z_\rho}
{\delta \bareta^+(y)\delta \bareta^+(x)\delta \eta^+(x)\delta \eta^+(y)}
\dd\bfr\dd\bfr',
\end{eqnarray*}
where $x=(t,\bfr)$, $y=(t,\bfr')$ and the right-hand side
is taken for zero external sources.
$Z_\rho$ is the generating function of the Green functions.
It is given by the expression $Z_\rho=\sum_{KL} \rho_{LK} Z_{KL}$,
with $\rho_{LK}$ the density matrix and
\begin{eqnarray*}
Z_{KL} &=& \langle K|S(\bareta^-,\eta^-)^{-1} S(\bareta^+,\eta^+)|L\rangle.
\end{eqnarray*}
Here, $S(\bareta^\pm,\eta^\pm)$ is the S-matrix in the
presence of fermionic external sources
$\bareta^\pm$ and $\eta^\pm$ and $Z_{KL}=\ee^{-iD} Z^0_{KL}$,
where $D=D_+-D_-$ and $D_\pm$ is the interaction term
$\int_{-\infty}^\infty H^\inter(t)\dd t$ where the fields are
replaced by functional derivatives with respect
to the external sources $\bareta_\pm,\eta_\pm$. Finally,
$Z^0_{KL}=\exp[-i\int \bfbareta(x)
G^0_0(x,y)\bfeta(y)\dd x\dd y]
N^0_{KL}$. In this equation,
$\bfeta$ is a two-dimensional vector with components
$\eta_+$ and $\eta_-$, $G^0_0(x,y)$ is the 2x2 matrix
\begin{eqnarray*}
\left( \begin{array}{cc}
    -i\langle 0 | T\big(\psi(x)\barpsi(y)\big)|0\rangle
       & -i\langle 0 | \barpsi(y)\psi(x)|0\rangle \\
   i\langle 0 | \psi(x)\barpsi(y)|0\rangle
        & -i\langle 0 |
\barT\big(\psi(x)\barpsi(y)\big)|0\rangle
         \end{array}\right),
\end{eqnarray*}
$\barT$ is the anti-time-ordering operator
and
$N^0_{KL}=\langle K|{:}\exp\big[ i\int
\bareta_d(x)\psi(x)+\barpsi(x)\eta_d(x)\dd x\big]{:}|L\rangle$,
with $\eta_d=\eta_+-\eta_-$.

To calculate $N^0_{KL}$, we must define 
the states $|K\rangle$ and $|L\rangle$.
The solution of the noninteracting Schr\"odinger
equation provides one-electron orbitals
$u_n(\bfr)$ with energy $\epsilon_n$.
An orbital is called a core orbital if it is filled
in all states $|K\rangle$ and $|L\rangle$,
otherwise, it is called a valence orbital.
The core orbitals are numbered from 1 to $C$,
the valence orbitals from 1 to $M$.
There are $C$ electrons in the core orbitals
and $N\le M$ electrons in the valence orbitals.
For example, in the ion Cr$^{3+}$,
there are $C=18$ core orbitals and
$N=3$ electrons in the $M=10$ orbitals
of the degenerate 3d shell.
The states are generated from the vacuum $|0\rangle$
by the action of creation operators $c^\dagger_n$
for core electrons and $v^\dagger_n$ for valence electrons
as
\begin{eqnarray*}
|K\rangle &=& v^\dagger_{i_N}\dots v^\dagger_{i_1}
             c^\dagger_C\dots c^\dagger_{1}|0\rangle,\\
|L\rangle &=& v^\dagger_{j_N}\dots v^\dagger_{j_1}
             c^\dagger_C\dots c^\dagger_{1}|0\rangle,
\end{eqnarray*}
where $i_k$, $j_k$ are valence orbitals
(i.e. integers taken in the set
$\{1,\dots,M\}$) ordered so that
$i_1<\dots<i_N$ and $j_1<\dots<j_N$.
A lengthy calculation \cite{BrouderKB} yields
\begin{eqnarray*}
N^0_{KL} &=& \prod_{k=1}^C (1+\bara_k\alpha_k)
\\&&\times
\exp\Big(\sum_{n=1}^M \frac{\partial^2}{\partial \alpha_n
      \partial \bara_n}\Big) \bara_{j_1} \alpha_{i_1}
      \dots \bara_{j_N} \alpha_{i_N},
\end{eqnarray*}
with $\bara_n=\int \bareta_d(x)u_n(x)\dd x$
and $\alpha_n=\int \baru_n(x)\eta_d(x)\dd x$,
where $u_n(x)=\ee^{-i\epsilon_n t} u_n(\bfr)$,
$\baru_n(x)=\ee^{i\epsilon_n t} \baru_n(\bfr)$
and $x=(t,\bfr)$. 

Notice that, for vanishing sources (i.e. when $\eta^\pm=\bareta^\pm=0$), 
$Z_{KL}=\delta_{K,L}$. To calculate $E(\rho)$, it
is also possible to use the Galitskii-Migdal formula \cite{Galitskii},
but our expression for $E(\rho)$ allows for a more
direct comparison with the standard equations
of the crystal field theory.

{\sl{Energy minimization}} --
The minimization of $E(\rho)$ with respect to $\rho$
is constrained by the fact that $\rho$ must be
Hermitian, nonnegative and with unit trace.
To remove the trace condition, we use the
fact that, for a general matrix $\rho$,
$Z_\rho=\tr(\rho)$ for vanishing sources.
So the trace condition is relaxed by writing the
energy as $E(\rho)=P(Z_\rho)/Z_\rho$.
The conditions of Hermiticity and nonnegativeness are
relaxed by writing $\rho=b^\dagger b$, where
$b$ is now unconstrained.
If we take the derivative of $E(\rho)=P(Z_\rho)/Z_\rho$
with respect to $b^\dagger$ and look for an extremum
where $\partial E(\rho)/\partial b^\dagger=0$ we obtain the
equation
$E(\rho)\partial Z_\rho/\partial b^\dagger=P(\partial Z_\rho/\partial b^\dagger)$ 
or, more explicitly
$E(\rho) \sum_{M} b_{KM} Z_{ML}=\sum_{M} b_{KM} P(Z_{ML})$.
For vanishing sources we have $Z_{ML}=\delta_{L,M}$ and this
equation becomes
$E(\rho)b=bP(Z)$, where $Z$ is the matrix $Z_{ML}$.

To solve this, we diagonalize $P(Z)$. This gives us a unitary matrix
$U$ and a diagonal matrix $z$ such that 
$E(\rho)b=b UzU^\dagger$, or
$E(\rho)bU=bU z$.

To determine the general solution of
this equation, we order the eigenvalues and eigenvectors
of $P(Z)$ and we call $e_0$ the smallest eigenvalue.
We put $R=bU$, and we assume that
the lowest eigenvalue is $d$-fold degenerate.
The general solution of $E(\rho)bU=bU z$
(i.e. $E(\rho)R=Rz$) giving the minimum energy is
$E(\rho)=e_0$ and $R_{ij}=0$ for $j>d$.
Then $b=RU^\dagger$ and $\rho=U R^\dagger R U^\dagger$,
where $R^\dagger R$ is an arbitrary Hermitian nonnegative
$d$-dimensional matrix.
At this stage, we can determine $R$ by considering the symmetry of the
system. The solution which gives the
same weight to each eigenstate (and then the
maximum symmetry) is $R^\dagger R=1_d$.
This determines the density matrix to be
\begin{eqnarray}
\rho_{KL} &=& \frac{1}{d} \sum_{M=1}^d U_{KM} U^\dagger_{LM},
\end{eqnarray}
where the factor $1/d$ was added to have $\tr(\rho)=1$.
The maximum symmetry can be broken by choosing another
$d$-dimensional matrix $R^\dagger R$.
For example, in a spherically symmetric system,
the matrix $R^\dagger R$ can be chosen to select
a specific spectroscopic term $^{2S+1}L$. 
In systems where the spin-orbit interaction is weak,
we can select a given spin state. This
can be useful to make self-consistent calculations
of low-spin and high-spin transition metal ions.

{\sl{The effective matrix}} --
It remains now to determine the effective matrix
$P(Z)$. We can obtain $P(Z)$ by noticing that
$Z_\rho=\sum_{KL} \rho_{LK} Z_{KL}$ implies
$Z_{KL}=\partial Z_\rho/\partial\rho_{LK}$.
Therefore, the equation for $Z_{KL}$ can be
obtained from an equation for $Z_\rho$ by
a derivation with respect to $\rho_{LK}$.

The equation for $Z_\rho$ is \cite{BrouderKB}
\begin{eqnarray}
\partial Z_\rho &=& \sum_{n=0}^\infty
\frac{(-i)^n}{n!} \sum (-1)^{|D\i{1'}^{n}|}
(D\i{1'}^{n} \partial W^1_\rho) (D\i{2'}^{n} Z_\rho),
\nonumber\\
\label{partialZrho}
\end{eqnarray} 
where $\partial$ is a functional derivative
with respect to an external source $\bareta^{\pm}(x)$
or $\eta^{\pm}(x)$.
In fact, we need an infinite system of equations
(the Green function hierarchy) obtained by taking
the functional derivative of equation (\ref{partialZrho}):
\begin{eqnarray}
d\partial Z_\rho &=& \sum_{n=0}^\infty
\frac{(-i)^n}{n!} \sum (-1)^{f}
(d\i1 D\i{1'}^{n} \partial W^1_\rho) (d\i2 D\i{2'}^{n} Z_\rho),
\nonumber\\
\label{dpartialZrho}
\end{eqnarray}
where $f=|D\i{1'}^{n}|+|d\i2||D\i{1'}^{n}| + |d\i2|$,
$d$ is a polynomial in the functional derivatives
with respect to an external source $\bareta^{\pm}(x)$
or $\eta^{\pm}(x)$ and all quantities are taken 
for vanishing sources (i.e.  $\bfbareta=\bfeta=0$). 
In the rest of the discussion,
we consider only the equation for $\partial Z_\rho$,
the other equations being obtained through further functional
derivatives, as for equation (\ref{dpartialZrho}).

Taking the partial derivative of equation
(\ref{partialZrho}) with respect to $\rho_{LK}$,
we obtain an equation for $\partial Z_{KL}$:
\begin{eqnarray}
\partial Z_{KL} &=& \sum_{n=0}^\infty
\frac{(-i)^n}{n!} \sum (-1)^{|D\i{1'}^{n}|}\Big(
(D\i{1'}^{n} \partial W^1_{KL}) (D\i{2'}^{n} Z_\rho)
\nonumber\\&&+ (D\i{1'}^{n} \partial W^1_\rho) (D\i{2'}^{n} Z_{KL}),
\label{partialZKL}
\end{eqnarray} 
where $W^1_{KL}=\partial W^1_\rho/\partial \rho_{LK}$.
For the calculation of the partial derivative of
$ W^1_\rho$ with respect to $ \rho_{LK}$,
it is necessary that a general value of $\rho$ be
used in $ W^1_\rho$. In particular, $\rho$ must
not be assumed Hermitian or with unit trace. This implies
the presence of terms such as $1/\tr(\rho)$  in
 $ W^1_\rho$. Once the partial derivative is calculated,
the condition $\tr(\rho)=1$ can be restored to simplify
$W^1_{KL}$.
Notice that $ W^1_\rho$ is not linear in $\rho$,
so $W^1_{KL}$ depends on $\rho$. This dependence 
implies that the solution of the system of equations
for $\rho$, $Z_\rho$ and $Z_{KL}$ must be solved
self-consistently.

We are interested in the energy splitting due to the
interaction. So we can remove the terms that are
proportional to $\delta_{KL}$ because they
just shift all eigenvalues by a constant term
and do not modify the eigenvectors.
To do that we write
$Z_{KL}=Y_{KL}+ (Z_\rho  + \barY )\delta_{KL}$,
with the boundary conditions $Y_{KL}=0$ and $\barY=0$
for vanishing external sources. Introducing this
into equation (\ref{partialZKL}), we obtain 
the following equation for $Y_{KL}$:
\begin{eqnarray}
\partial Y_{KL} &=& \sum_{n=0}^\infty
\frac{(-i)^n}{n!} \sum (-1)^{|D\i{1'}^{n}|}\Big(
(D\i{1'}^{n} \partial V_{KL}) (D\i{2'}^{n} Z_\rho)
\nonumber\\&&+ (D\i{1'}^{n} \partial W^1_\rho) (D\i{2'}^{n} Y_{KL})\Big),
\label{partialYKL}
\end{eqnarray} 
where 
\begin{eqnarray*}
V_{KL} &=& W^1_{KL} - \frac{\delta_{KL}}{\dim} \sum_K  W^1_{KK},
\end{eqnarray*}
with $\dim$ the dimension of the matrix $W^1_{KL}$.

Finally, the algorithm for the calculation of the density matrix
$\rho$ and the Green functions $d Z_\rho$ is the following:
(i) choose an initial density matrix $\rho$, (ii) calculate
$W^1_\rho$ and $V_{KL}$, (iii) use equation (\ref{dpartialZrho})
to calculate the $d Z_\rho$ you need for the calculation of the energy,
(iv) calculate $dY_{KL}$ for a given value of $dZ_\rho$ and
$V_{KL}$, (v) diagonalize $P(Y_{KL})$, (vi) deduce a new density
matrix $\rhonew$, (vii) compare $\rhonew$ and $\rho$, if convergence
is not reached, go back to (ii) with a new $\rho$.
Remark that steps (iii) and (iv) require the solution
of an infinite system of equations. This system must
be broken by some closure approximation of the Green function
hierarchy and the resulting $d Z_\rho$ and  $dY_{KL}$
are thus approximate. Notice also that
another self-consistency loop can be introduced in the algorithm
to calculate a self-consistent external potential, as was
described in \cite{BrouderKB0}. This will change the
orbitals $u_n(x)$.

{\sl{The crystal field equations}} --
We must show that the present approach contains the standard 
crystal field equations as a first-order approximation.
The functions we need to calculate $P(Y_{KL})$ are
$\partial\partial' Y_{KL}$ with $\partial'=\delta/\delta\eta^+(y)$
and  $\partial=\delta/\delta\bareta^+(x)$ and
$\partial_1\partial_2\partial_3\partial_4 Y_{KL}$
with
$\partial_4=\delta/\delta\eta^+(y)$,
$\partial_3=\delta/\delta\eta^+(x)$,
$\partial_2=\delta/\delta\bareta^+(x)$ and
$\partial_1=\delta/\delta\bareta^+(y)$.
If we solve iteratively equations (\ref{dpartialZrho})
and (\ref{partialYKL}) for $n=0$ we obtain
\begin{eqnarray}
\partial\partial' Y_{KL} &=& \partial\partial' V_{KL},
\label{crystalfield2}
\\
\partial_1\partial_2\partial_3\partial_4 Y_{KL} &=&
\partial_1\partial_2\partial_3\partial_4 V_{KL}
- (\partial_2\partial_4 V_{KL})(\partial_1\partial_3 Z_\rho)
\nonumber\\&&\hspace*{-13mm}
+ (\partial_1\partial_4 V_{KL})(\partial_2\partial_3 Z_\rho)
- (\partial_2\partial_4 W^1_\rho)(\partial_1\partial_3 Y_{KL})
\nonumber\\&&\hspace*{-13mm}
+ (\partial_1\partial_4 W^1_\rho)(\partial_2\partial_3 Y_{KL}).
\label{crystalfield4}
\end{eqnarray}
A lengthy calculation shows that $P(Y_{KL})$ calculated
from  (\ref{crystalfield2}) and  (\ref{crystalfield4}) gives the
same matrix as the standard crystal field theory.

{\sl{The simplest example}} --
As an example, we consider the case of one electron (N=1) 
that can occupy two orbitals (M=2).  The density matrix
is then of dimension 2.  The matrix $V_{KL}$ takes the form
\begin{eqnarray*}
V_{KL} &=& \left( \begin{array}{cc}
    (\bara_1\alpha_1 - \bara_2\alpha_2)/2 & \bara_2\alpha_1 \\
    \bara_1\alpha_2  & (\bara_2\alpha_2 - \bara_1\alpha_1)/2 \end{array}\right)
\\&&
+ \bara_1\alpha_1 \bara_2\alpha_2
\left( \begin{array}{cc}
    (\rho_{11}-\rho_{22})/2 & \rho_{12} \\
   \rho_{21} & (\rho_{22}-\rho_{11})/2 \end{array}\right).
\end{eqnarray*}
In particular, if the system is degenerate ($\rho_{11}=\rho_{22}=1/2,
\rho_{12}=\rho_{21}=0$), the second term is zero and $V_{KL}$ has
only a term linear in $\bara$ and $\alpha$.

We presented a first-principles self-consistent
many-body approach to crystal field theory.
Our approach contains by construction the
many-body Green function theory as the case $N=M$
and it contains the crystal field theory as a first
order approximation. Thus, it is expected to provide good
results in situations that are intermediate between the
range of validity of these two limits, such as in 
strongly-interacting electron systems.

This method can be used
more generally to calculate the splitting,
due to interaction, of levels that are degenerate
in the noninteracting system, for any
quantum field theory. A interesting example
is the application of this approach to 
the quantum electrodynamics of atoms,
where its self-consistency  should make it
competitive with respect to the effective
Hamiltonian methods proposed by
Shabaev \cite{Shabaev} and Lindgren
\cite{Lindgren1}.

\begin{acknowledgments}
I am very grateful to S. Di Matteo for mentioning 
the important reference \cite{Esterling}.
This is IPGP contribution \#0000.
\end{acknowledgments}

%
% ****** End of file apssamp.tex ******
\end{document}